\documentclass[dvipsnames, letterpaper, 10 pt, conference]{ieeeconf}  % Comment this line out
\IEEEoverridecommandlockouts                              % This command is only
\usepackage{graphicx}
\usepackage{pgfplots}
\pgfplotsset{compat=1.18}

\usepackage{mathptmx} % assumes new font selection scheme installed
\usepackage{times} % assumes new font selection scheme installed
\usepackage{amsmath} % assumes amsmath package installed
\usepackage{amssymb}  % assumes amsmath package installed
\usepackage{tikz}

\usepackage[american,siunitx]{circuitikz}
\usetikzlibrary{shapes.geometric, decorations.pathreplacing,calc,positioning,arrows.meta}

\tikzstyle{LTI} = [rectangle, minimum width=2cm, minimum height=1cm,text centered, draw=black]
\tikzstyle{a} = [rectangle, minimum width=1cm, minimum height=1cm,text centered, draw=black]
\tikzstyle{arrow} = [thick,->,>=stealth, line cap = round]
\tikzstyle{box} = [rectangle, minimum width=1cm, minimum height=1cm,text centered, draw=black]

\title{\LARGE \bf
   A memristive model of spatio-temporal excitability
}

\author{Thomas S.\,J. Burger, Amir Shahhosseini, Rodolphe Sepulchre% <-this % stops a space
\thanks{KU Leuven, Department of Electrical Engineering (ESAT), STADIUS Center for
Dynamical Systems, Signal Processing and Data Analytics,
Kasteelpark Arenberg 10, 3001 Leuven, Belgium
        {\tt\small thomas.burger@kuleuven.be}}%
}

\begin{document}

\maketitle
\thispagestyle{empty}
\pagestyle{empty}

%%%%%%%%%%%%%%%%%%%%%%%%%%%%%%%%%%%%%%%%%%%%%%%%%%%%%%%%%%%%%%%%%%%%%%%%%%%%%%%%
\begin{abstract}
This paper introduces a model of excitability that unifies the mechanism of an important neuronal property both in time and in space. As a starting point, we revisit both a key model of temporal excitability, proposed by Hodgkin and Huxley, and a key model of spatial excitability, proposed by Amari. We then propose a novel model that captures the temporal and spatial properties of both models. Our aim is to regard neuronal excitability as a property \emph{across scales}, and to explore the benefits of modeling excitability with one and the same mechanism, whether at the cellular or the population level.
   %Studying the behavior of neuronal circuits over their full range of temporal and spatial scales is challenging,
   %as these scales span orders of magnitudes,
   %and these levels of activity generally interact with one another.
   %This means that relating macroscopic behaviors to microscopic, cellular properties is difficult.
   %Therefore, if one wants to design and analyze neuromorphic circuits operating over similar scales,
   %deciding which neuronal properties to emulate is unclear.
   %To address these questions, we introduce a model
   %whose behavior across scales of both time and space is easily controlled by its physical circuit elements,
   %namely its conductances.
   %We show that this model unifies the temporal excitability of biophysical neuron models
   %and the spatial excitability of mean-field models.
   %We dub this unified excitability property spatio-temporal %excitability.
\end{abstract}

%%%%%%%%%%%%%%%%%%%%%%%%%%%%%%%%%%%%%%%%%%%%%%%%%%%%%%%%%%%%%%%%%%%%%%%%%%%%%%%%
\section{INTRODUCTION}

%\tsjb{
%TODO:
%\begin{itemize}
%   \item Let Amir \& Robin know when the ArXiV pre-print is online so they can reference it.
%   \item Should the different models and equations be wrapped in Definition statements?
%\end{itemize}
%}

%\rs{Test this colour.}

Neuronal circuits function on spatial and temporal scales that span several orders of magnitude,
from milliseconds to years, and from micrometers to dozens of centimeters.
As just one example, neuron activity in the brain displays synchrony across multiple temporal and spatial scales,
which has implications for a range of functions~\cite{kahanaCognitiveCorrelatesHuman2006}.

Modeling excitability has been a key focus of mathematical modeling in neuroscience since the seminal work of Hodgkin and Huxley in 1952 (see \cite{hodgkinQuantitativeDescriptionMembrane1952} and the  recent textbooks \cite{keener2009mathematical,izhikevich2007dynamical,ermentrout2010foundations}). The tradition has been, however, to adopt different modeling principles to capture excitability at different scales. The  biophysical principles of Hodgkin and Huxley have been broadly adopted to model excitability at the level of a single neuron or for small circuits involving just a few neurons (for instance \cite{buchholtz1992mathematical}). Instead, modeling excitability at the level of neuronal populations has been addressed with mean-field approaches (e.g.\ the model of Wilson and Cowan \cite{wilsonExcitatoryInhibitoryInteractions1972} and the model of Amari \cite{amariDynamicsPatternFormation1977}). While cellular and population models of excitability share a number of qualitative properties, they are nevertheless of a different nature. Both cellular and mean-field models have been successful at describing neuronal behaviors at distinct scales, but the separate nature of the models has made it difficult to study excitability {\it across} scales, that is, to explore how  cellular excitability mechanisms shape and modulate population excitability mechanisms. Attempts have been made to 
derive the macroscopic behavior  from single-cell properties (see e.g.\ \cite{coombesNextGenerationNeural2019}),
but at the price of starting from abstract neuron models that are difficult to relate to biophysical modeling principles.

Previous work by the authors, in particular \cite{drion2018switchable} and the recent PhD dissertation \cite{burger_thesis} illustrate the shortcomings of modeling cellular and population activity with different modeling principles. In \cite{drion2018switchable}, the authors show how neuromodulators can shape spatio-temporal activity (``brain states'') by modulating the intrinsic properties of the neurons rather than the synaptic neural connections. In \cite{burger_thesis}, the author demonstrates how particular properties at the cellular scale determine the robustness of the mean-field PING rhythm at the population scale. In both examples, a population phenomenon is regulated at the cellular scale, and the only way to study the mechanism ``across scales'' is to simulate the entire population at the neuronal scale, which is cumbersome and not scalable. 

Motivated by those shortcomings, we aim at developing new population models that will make it easier to study neuronal behaviors across scales. As a first step in that direction, we propose a spatio-temporal model that unifies the temporal model of Hodgkin and Huxley and the spatial model of Amari. 

Our angle of attack is to retain the key biophysical modeling principles of Hodgkin and Huxley, namely the circuit representation of each neuron as the parallel connection of a capacitive element with a bank of voltage-gated current sources. In the original model of Hodgkin and Huxley, the current sources are {\it memristive}, that is, they obey an Ohmic law but with conductances that have memory, that is, depend on the history of the voltage. In the present paper, such conductances with memory are called {\it memductances}, following the terminology of \cite{chuaMemristiveDevicesSystems1976}. The proposed standpoint is that this memristive modelling principle is valid both at the cellular and the population scale, and that only the model of the memductance differs. We then exploit the close analogy between ion channels (i.e.\ {\emph{internal} memductances) and synapses (i.e.\ \emph{external} memductances) to model their temporal and spatial voltage dependence.

We stress the advantages of a simple circuit model that concentrates all the temporal and spatial modeling complexity in the spatio-temporal voltage dependence of the memductances. Simplified models preserve the same physical circuit structure but use simplified representations of the memductances. Temporal models spatially average the spatial dependence, while spatial models temporally average the voltage dependence. We argue that such models can be simulated, analyzed, and designed \emph{at scale} by modulating the model resolution of the memductances only. 

The paper is structured as follows.
In Section \ref{sec:temp-excite}, we review the Hodgkin Huxley model of  temporal excitability. Section \ref{sec:space-excite} revisits Amari's model of lateral inhibition, which we regard as spatial excitability.
The unified memristive model is introduced in Section \ref{sec:memristive-model}.
We show that simple models of the memductances capture both temporal and spatial excitability, either separately or together.

%%%%%%%%%%%%%%%%%%%%%%%%%%%%%%%%%%%%%%%%%%%%%%%%%%%%%%%%%%%%%%%%%%%%%%%%%%%%%%%%
\section{TEMPORAL EXCITABILITY}
\label{sec:temp-excite}

%\tsjb{
%\begin{itemize}
%   \item Introduce the HH model
%   \item Give the circuit diagram
%   \item Point out that it is a capacitor in parallel with several nonlinear resistors.
%   \item Demonstrate the excitability phenomenon, that is, the existence of a threshold.
%\end{itemize}
%}

The biophysical foundation of neuroscience,
and therefore of our understanding of excitable behaviors,
is the work of Hodgkin and Huxley,
culminating in \cite{hodgkinQuantitativeDescriptionMembrane1952}.

The key modeling principle is that a neuron can be represented by an RC circuit in parallel with a bank of voltage-gated current sources. Temporal excitability only requires two such sources, as shown  in Equation \ref{eq:HH-model}.
The capacitor models the membrane of the neuron,
because there is a voltage difference between the inside of the cell and the extracellular medium.
This voltage changes as ions flow in and out through the cell membrane.
The flow of each ion is controlled by a specific ion channel, modeled as an Ohmic current whose conductance is voltage-dependent due to the voltage-dependent opening
and closing of channels at the microscopic level.

In Hodgkin and Huxley's model, only two ion channels are present: sodium (Na) and potassium (K).
The remaining channels are gathered in a passive `leak' current $i_l$. The applied current $i_\text{app}$ models ``external'' currents, for instance resulting from synaptic connections to other neurons.
The equations of the Hodgkin-Huxley model are:
\begin{equation}
   \begin{aligned}
      C \dot{v} &= - \left( i_l + i_\text{Na} + i_\text{K} \right) + i_\text{app} \\
      i_l &= g_l \left( v - E_l \right) \\ 
      i_\text{Na} &= g_\text{Na}(t) \left( v - E_\text{Na} \right) \\ 
      i_\text{K} &= g_\text{K}(t) \left( v - E_\text{K} \right). \\ 
   \end{aligned}
   \label{eq:HH-model}
\end{equation}
Here, $v$ is the voltage over the cell membrane,
$C$ is the capacitance of the membrane, 
$g_k$ denotes the memductance of the current flowing through the ion channel of type $k$,
and $E_k$ is the equilibrium (or Nernst) potential of this channel.

%\begin{figure}
%   \centering
%   \begin{circuitikz}[american]
%      \draw (0,0) to[capacitor=$C$, v=$v$, voltage/shift=-6.5, voltage/american label distance=3.5, i>_=$i_l + i_m$] (0,4) to[short, -*] (2,4) to[R=$R_l$] (2,2) to[battery2=$E_l$, i^>=$i_l$] (2,0) -- (0,0);
%      \draw (2,4) to[short, -*] (4,4) to[vR=$R_{\text{Na}}$] (4,2) to[battery2=$E_\text{Na}$, i^>=$i_\text{Na}$] (4,0) to[short, *-*, i=$i_m$] (2,0);
%      \draw (4,4) to[short, -*] (6,4) to[vR=$R_{\text{K}}$] (6,2) to[battery2=$E_\text{K}$, i^>=$i_\text{K}$] (6,0) to[short, *-*] (4,0);
%   \end{circuitikz}
%   \caption{
%      The circuit diagram of the Hodgkin-Huxley model.
%   $R_l = 1 / g_l$, $R_\text{Na} = 1 / g_\text{Na}$, $R_\text{K} = 1 / g_K$.
%   $R_\text{Na}$ and $R_\text{K}$ are time-varying and voltage dependent,
%   giving rise to temporal excitability.
%}
%   \label{fig:conductance-model}
%\end{figure}

The temporal voltage-dependence of  the ionic memductances of sodium and potassium is modeled through additional state variables that model the activation and inactivation of the current. In the Hodgkin-Huxley model, the ionic memductance of ion $k$ is modeled by
\begin{equation}
      g_k(t, v) = \bar{g} m_k^a h_k^b,
      \label{eq:HH_conductance}
\end{equation}
for some $a, b \in \mathbb{N}$ and a maximal conductance $\bar{g}_k > 0$.
Here, $m$ and $h$ both have first-order dynamics
\begin{equation}
   \dot{x} = \left( x_\infty(v) - x \right) / \tau_x(v),
   \label{eq:HH_gating_var}
\end{equation}
where $x_\infty(v) \in [0,1]$ is a sigmoidal function,
and $\tau_x(v)$ is either bell-shaped or U-shaped,
depending on the ion in question.
$m$ is called an activation variable because $m_\infty(v)$ increases monotonically with $v$ (that is, it models the channel opening as $v$ increases),
and $h$ is called an inactivation variable as $h_\infty(v)$ decreases monotonically (it models the channel closing as $v$ increases).
For the sodium channel in the Hodgkin-Huxley model $a = 3$ and $b = 1$
and for the potassium channel $a = 4$ and $b = 0$.

The feedback structure of the model is sketched in Figure \ref{fig:HH-block-diagram}.
First, the voltage goes through a nonlinearity $x_\infty$ for $x = m,n,h$.
The resulting quantity goes through a nonlinear temporal convolution with an exponential kernel $e^{-t / \tau_x(v)}$
to give the gating variables.
These two steps capture Equation \ref{eq:HH_gating_var}.
Then, these gating variables go through polynomials that give the ionic currents,
which go through the $RC$ circuit to determine $v$.

\begin{figure}
   \centering
   \includegraphics[width=\columnwidth]{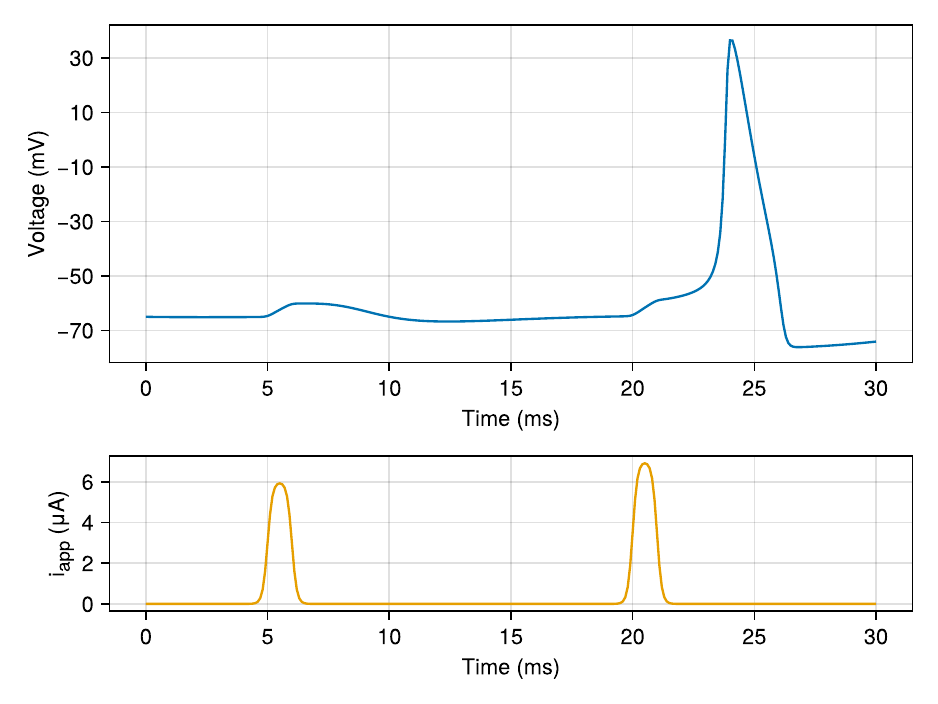}
   \caption{
      Temporal excitability in the Hodgkin-Huxley model.
      A small difference in $i_\text{app}$ (bottom) can cause a large difference in the voltage response (top).
   }
   \label{fig:HH-excitability}
\end{figure}

The Na and K ion channels together are sufficient to generate temporally excitable behaviors:
a stereotyped, brief large-amplitude response when the input stimulus exceeds some threshold in frequency or amplitude,
as illustrated in Figure \ref{fig:HH-excitability}.
The fast activation of the sodium current
provides positive feedback on $v$, causing it to act as a switch.
This fast switching behavior is captured in the model by $m$,
specifically through $m_\infty(v)$ increasing monotonically and $\tau_m(v)$ being small compared to $\tau_n$ and $\tau_h$, together with $E_\text{Na}$ being much bigger than the resting potential of the neuron $v_\text{rest}$ so that $i_\text{Na} < 0$
for normal voltage values of the neuron.
But this switch is then turned off through the much slower inactivation of sodium (modeled by $h$) and activation of potassium (modeled by $n$),
which provide negative feedback on $v$.
This motif of fast positive feedback paired with slower negative feedback for global stability
creates a region where the model is ultra-sensitive: 
this is the threshold, which lies at the heart of excitable behaviors in time.

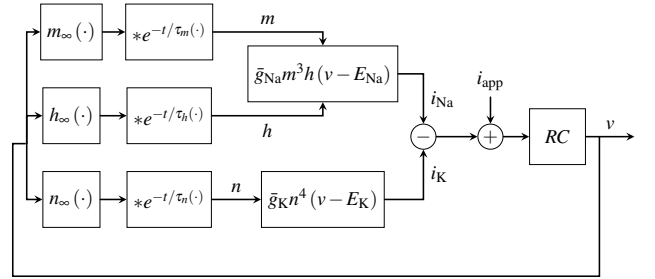
\begin{figure}
\centering
\resizebox{\columnwidth}{!}{
\begin{tikzpicture}[node distance=1.5cm]
	% Nodes
	\node (start) [inner sep=0pt, outer sep=0pt] at (0,0) {};
	\node (sigma_m) at ($(start) + (0.8cm,2.0cm)$) [box] {$m_\infty \left( \cdot \right)$};
	\node (conf_m) [a, right of=sigma_m, node distance=1.75cm] {$* e^{-t / \tau_m( \cdot )}$};
	\node (sigma_h) [box, below of=sigma_m] {$h_\infty \left( \cdot \right)$};
	\node (conf_h) [a, right of=sigma_h, node distance=1.75cm] {$* e^{-t / \tau_h( \cdot )}$};
	\node (Na_poly) at ($(conf_h)!0.5!(conf_m)+(0:2.75cm)$) [box] {$\bar{g}_\text{Na} m^3 h \left(v - E_\text{Na} \right)$};
	\node (sigma_n) at ($(start) + (0.8cm,-1.0cm)$) [box] {$n_\infty \left( \cdot \right)$};
	\node (conf_n) [a, right of=sigma_n, node distance=1.75cm] {$* e^{-t / \tau_n( \cdot )}$};
	\node (K_poly) at ($(conf_n) + (0:2.75cm)$) [box] {$\bar{g}_\text{K} n^4 \left(v - E_\text{K} \right)$};
	\node (minus) at ($(Na_poly)!0.5!(K_poly)+(0:1.8cm)$) [circle, minimum size=4.5mm, draw, inner sep=0pt] {$-$};
	\node (plus) at ($(minus) + (1.2cm,0)$) [circle, minimum size=4.5mm, draw, inner sep=0pt] {$+$};
	\node(i_app) at ($(plus)+(0,1cm)$) {$i_\text{app}$};
	\node(RC) at ($(plus)+(1.2cm,0)$) [box] {$RC$};
	\node(end) at ($(RC)+(1.5cm,0)$) {};

    % Arrows
    \draw [arrow] (start.center) |- (sigma_m);
    \draw [arrow] (sigma_m) --  (conf_m);
    \draw [arrow] (conf_m) -| (Na_poly) node[near start, above] {$m$};
    \draw [arrow] (Na_poly) -|  node[near end, right] {$i_\text{Na}$} (minus);
    \draw [arrow] (start.center) |- (sigma_h);
    \draw [arrow] (sigma_h) -- (conf_h);
    \draw [arrow] (conf_h) -| (Na_poly) node[near start, below] {$h$};
    \draw [arrow] (start.center) |- (sigma_n);
    \draw [arrow] (sigma_n) -- (conf_n);
    \draw [arrow] (conf_n) -- (K_poly) node[midway, above] {$n$};
    \draw [arrow] (K_poly) -|  node[near end, right] {$i_\text{K}$} (minus);
    %
    %\draw [arrow] ($(RC)!0.5!(end)$) -- ++(0,2cm) -| ($(Na_poly.north) + (0.5cm,0)$);
    %\draw [arrow] ($(RC)!0.5!(end)$) -- ++(0,-2cm) -| ($(K_poly.south) + (0.5cm,0)$);
    %
    \draw [arrow] (minus) -- (plus);
    \draw [arrow] (i_app) -- (plus);
    \draw [arrow] (plus) -- (RC);
    \draw [arrow] (RC) -- node[above] {$v$} (end) ;
    \draw [thick] ($(RC)!0.5!(end)$)
	-- ++(0,-2.5)
	-- ++(-10.5,0)
	|- (start);

\end{tikzpicture}
}
\caption{Block diagram of the mixed temporal monotone structure of the Hodgkin-Huxley model.}
\label{fig:HH-block-diagram}
\end{figure}

%%%%%%%%%%%%%%%%%%%%%%%%%%%%%%%%%%%%%%%%%%%%%%%%%%%%%%%%%%%%%%%%%%%%%%%%%%%%%%%%
\section{SPATIAL EXCITABILITY}
\label{sec:space-excite}

%\tsjb{
%\begin{itemize}
%   \item Introduce Amari's model.
%   \item Point out that excitability in space is a localization response.
%   \item Point out the analogy of the (fast Na / slow K) in HH and the (short E / long I)  in Amari's model;
 %     the combination gives one excitability.
 %  \item QUESTION: I am not very convinced how accurate short E / long I is for real neural circuits:
  %    should we touch upon this assumption, defend it, or point out it's unrealistic?
%\end{itemize}
%}

Information in the brain is routed and shared through many synaptic stages and brain areas.
In the brain, the spatial extent of activity is mediated through a zoo of neurons,
which can be broadly characterized into two classes: excitatory (E) and inhibitory (I).
Excitatory neurons promote further neuronal activity,
whilst inhibitory neurons usually diminish it.
Excitation and inhibition are spatially structured,
with E and I operating in a balanced regime.
This can lead to interesting spatial patterns, such as oscillations, in the neuronal activity.
Two foundational models of this are \cite{wilsonExcitatoryInhibitoryInteractions1972} and \cite{amariDynamicsPatternFormation1977}.

In Amari's model \cite{amariDynamicsPatternFormation1977}, spatial interactions are modeled as a simple spatial convolution operator.
The spatio-temporal membrane potential is now $u(x,t)$.
The output (which models the average firing rate) is given by $f(u(x,t))$ for some monotonically non-decreasing $f$.
We use $f(x) = 1 / \left( 1 + \exp\left(-5 (u - \theta) \right) \right)$,
where $\theta$ is the firing threshold.
Then the neural field evolves according to
\begin{equation}
   \tau \dot{u} = -u + \int_\Omega w(x - y) f(u(y,t)) \, dy + i_\text{app},
   \label{eq:amari_model}
\end{equation}
where $w(x-y)$ represents the average connection strength from neurons at $y$ to neurons at $x$,
$\Omega$ is the region of the neural field,
and $\tau$ is the membrane time constant.

\begin{figure}
   \centering
   \includegraphics[width=0.8\columnwidth]{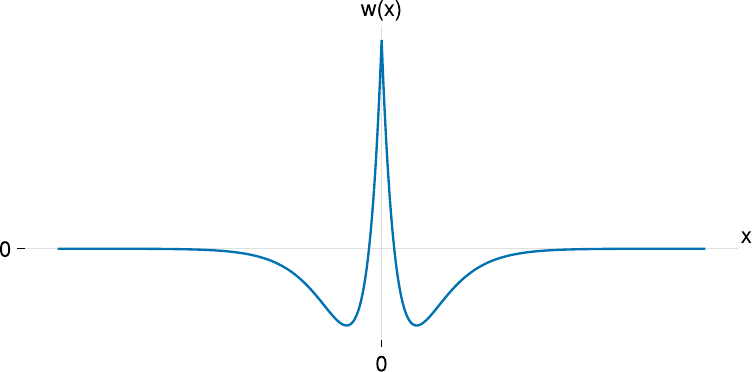}
   \caption{Sketch of $w(x)$ as used in Amari's model.}
   \label{fig:amari-w-kernel}
\end{figure}

The spatial interactions between neurons in the field are therefore determined by $w(x)$.
It is sketched in Figure \ref{fig:amari-w-kernel}.
Excitatory connections are those connections that lead to an increase in $\dot{v}$,
whereas inhibitory connections decrease $\dot{v}$.
The salient features of the interaction kernel are that the excitation ($w(x) > 0$) is short-range,
whereas inhibition ($w(x) < 0$) is longer range.

When the kernel is defined as the difference of exponentials $w(x) = e^{- |x| / \sigma_E} - e^{- |x| / \sigma_I}$,
the feedback structure of Amari's model can be represented as in Figure \ref{fig:amari-block-diagram}.
The averaged potential $u$ goes through $f$, 
the result is convolved in space with two kernels
with different parameters $\sigma$, 
defining the length scale of the interactions.
We require $\sigma_E < \sigma_I$ to preserve the difference in length scales between the E and I interactions.
Then, $i_E - i_I + i_\text{app}$ goes through the $RC$ block to produce $u$.
The reader will note the analogy with the block diagram of Hodgkin-Huxley model (Figure \ref{fig:HH-block-diagram}):
in both models, the voltage-dependent currents create a mixed-feedback model, that is, the feedback controller is the difference of two currents. The models only differ in how they model the voltage dependence of the currents: temporal nonlinear convolution with an exponential kernel in Hodgkin-Huxley model and spatial convolution with an exponential kernel  in Amari's model. 

As in the Hodgkin-Huxley model, the mixed-feedback structure of Amari's model is key to its spatial excitability:   when the neural mass exceeds a threshold,
the feedback mechanism leads to a spatially localized `event' or `spike'.  
Namely, the non-equilibrium behavior consists of one or more regions with a certain width, 
determined by the spatial kernel $w$,
that switch `on', thereby inhibiting the rest of the field.

This is demonstrated in Figure \ref{fig:amari_spatial_excitability},
where we have plotted the steady-state after relaxation $u^{ss}(x)$
in response to two Gaussian input pulses $i_\text{app}(x,t)$, both wide in space and time.
The first pulse does not cause $v$ to exceed its threshold, and the response is small.
The second pulse triggers an excitation in $u$ which is much narrower in space than the pulse, and persistent in time.
Many more complicated static and dynamical patterns can be supported, such as oscillations and traveling waves \cite{amariDynamicsPatternFormation1977}.
Spatial excitability is the system's ability to support these patterns reliably based on the statistics of the excitatory and inhibitory connections within the circuit.

\begin{figure}
   \centering
   \includegraphics[width=\columnwidth]{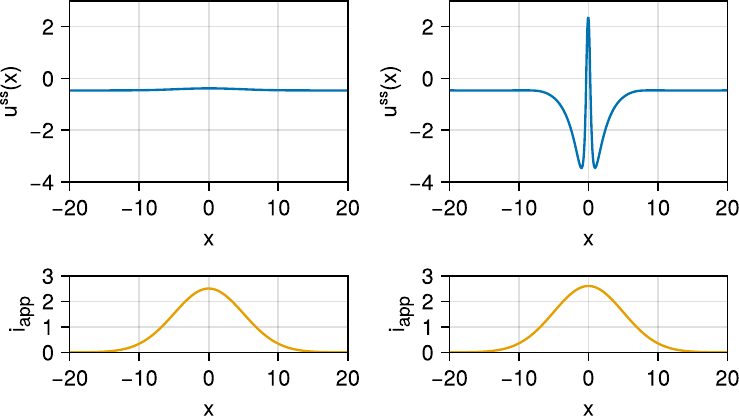}
   \caption{
      Spatial excitability in Amari's model.
      Left: the subthreshold steady-state voltage response $u^{ss}(x)$ (top)
      to Gaussian input current (bottom).
      Right: The superthreshold steady-state voltage response $u^{ss}(x)$ (top)
      to a Gaussian input current (bottom).
      The first applied pulse does not lead to spatial excitation,
      whereas the second applied pulse does,
   }
   \label{fig:amari_spatial_excitability}
\end{figure}

\begin{figure}
    \centering
    \resizebox{\columnwidth}{!}{
       \begin{tikzpicture}[node distance=1.5cm]
	% Nodes
	\node (start) [inner sep=0pt, outer sep=0pt] at (0,0) {};
	\node (nonlinearity) at ($(start) + (0.0cm,0)$) [box] {$f(\cdot)$};
	\node (right_nonlinearity) at ($(nonlinearity) + (1.0cm,0)$) {};
	\node (conf_s) at ($(start)+(20:2.5cm)$) [box] {$* \, e^{- |x| / \sigma_E}$};

	\node (conf_l) at ($(start)+(-20:2.5cm)$) [box] {$* \, e^{- |x| / \sigma_I}$};
	\node (E-I_plus) at ($(conf_s)!0.5!(conf_l)+(0:1.8cm)$) [circle, minimum size=4.5mm, draw, inner sep=0pt] {$+$};
	\node (plus) [right of=E-I_plus, circle, minimum size=4.5mm, draw, inner sep=0pt] {$+$};
	\node(i_app) at ($(plus)+(0,1cm)$) {$i_\text{app}$};
	\node(RC) at ($(plus)+(1.5cm,0)$) [box] {$RC$};
	\node(end) at ($(RC)+(2cm,0)$) {};

    % Arrows
	\draw [thick] (nonlinearity) -- (right_nonlinearity.center);
    \draw [arrow] (right_nonlinearity.center) |- (conf_s);
    \draw [arrow] (conf_s) -|  node[near start, above] {$i_E$} (E-I_plus);
    \draw [arrow] (right_nonlinearity.center) |- (conf_l);
    \draw [arrow] (conf_l) -|  node[near start, above] {$i_I$} node[very near end, right] {$-$} (E-I_plus);
    \draw [arrow] (E-I_plus) -- (plus);
    \draw [arrow] (i_app) -- (plus);
    \draw [arrow] (plus) -- (RC);
    \draw [arrow] (RC) -- node[above] {$u$} (end) ;
    \draw [arrow] ($(RC)!0.5!(end)$)
	-- ++(0,-1.75)
	-- ++(-9.0,0)
	|- (nonlinearity);
    \end{tikzpicture}
    }
    \caption{
        Block diagram of the mixed spatial monotone structure of Amari's model.
        We require $\sigma_E < \sigma_I$ to ensure that the excitatory interactions given by $i_E$
        operate on shorter length scales than the inhibitory interactions given by $i_I$.
    }
    \label{fig:amari-block-diagram}
\end{figure}
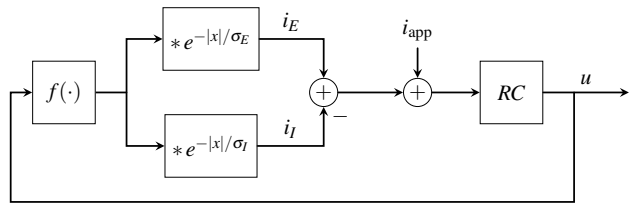

\section{MEMRISTIVE MODELING}
\label{sec:memristive-model}

%\tsjb{
%   \begin{itemize}
%   \item Explain what me mean by a memristive model.
%   We retain two elements from the Hodgkin-Huxley model:
%   \begin{itemize}
%       \item Each current is memristive (a conductance multiplied by the membrane voltage shifted by a battery voltage,
 %      \item Details of the conductance model are irrelevant, only the mixed feedback from the block diagrams is relevant
 %  \end{itemize}
  % \item Propose that each conductance could depend on both time and space,
  % and retain the mixed feedback structure of (fast / short range) inward current and (slow / long range) outward current.
  % \item The architecture of the memristive model is $\dot v = -g_l (v-v_l) - g_e (v-v_e) - g_i (v-v_i)$
  % \item The spatio-temporal complexity goes into how to model the $g_e$ and $g_i$.
   %\item Once this is explained, we discuss temporal excitability by having a simplified model for $g_e$ and $g_i$
   %\item Item for the simplification to a spatial excitability model.
%   \end{itemize}
%}
Based on the mixed-feedback analogy between the temporal excitability of the Hodgkin-Huxley model and the spatial excitability of Amari's model, we introduce a unified model that captures  both temporal and spatial excitability.
We retain the two important elements from the Hodgkin-Huxley model (Equation \ref{eq:HH-model}).
First, each current is memristive \cite{chuaMemristiveDevicesSystems1976}: the value of the current at time $t$ and position $x$  equals the membrane voltage at time $t$ and position $x$, multiplied by a memductance that depends on the entire neural field, or more precisely the entire past of the entire spatial neural field. Each memristive current is possibly in series with a constant voltage battery, resulting in a current source of the form $i=g(v-v_0)$ for some constant $v_0$.

Secondly, we retain the mixed feedback motif shown in Figure \ref{fig:HH-block-diagram} as the crucial mechanism for excitable behaviors.

Mimicking the architecture of the Hodgkin-Huxley model, 
we include one single inward current, labeled $e$, and one single outward current, labeled $i$.
The excitatory $e$ current models the fast Na current in Hodgkin-Huxley and / or the short ranged current in Amari,
while the inhibitory $i$ current models the slow K current and / or the long ranged current.

The voltage dynamics are that of a nonlinear RC circuit,
with the addition of a spatial dependence:
\begin{equation}
\begin{aligned}
    C \dot{v} = &- i_\text{tot} \\
                    = &-g_l v -
                        g_e(x, t)( v - E_e )
                        - g_i(x, t)( v - E_i )
    \label{eq:CNN_v_dot}
\end{aligned}
\end{equation}
The entire spatio-temporal complexity of the model goes into the modeling of the $g_e$ and $g_i$ memductances. Memductances must be nonlinear, because they modulate the current between zero (completely closed channels) and a maximum conductance $\bar g$ (completely open channels). The temporal dependence of the memductance must be ``fast'' or ``slow'', which we capture by a simple temporal convolution operation with decaying exponentials characterized by a specific time scale.  Likewise, the spatial dependence of the memductance must be ``short-range'' or ``long-range'', which we capture with a simple spatial convolution with decaying exponentials characterized by a specific spatial scale. 

The core observation is that the memductance of each current can be modeled as a nonlinear convolution operator,
typically in the form of a Convolution Neural Network (CNN) model.
The complexity of a particular instantiation of the model is only determined by the complexity of this {\it feedforward} neural network operator. In the next sections we will illustrate that simple models of the memductance suffice to capture temporal and spatial excitability.

\subsection{A model of temporal excitability}
\label{subsec:canonical_temporal_excitability}

In Section \ref{sec:temp-excite},
we discussed how temporal excitability is the result of fast positive feedback on the membrane potential
followed up with slower negative feedback.
How difficult is it to model such memristive properties with a CNN model?

To obtain temporal excitability, we restrict $g_e$ and $g_i$ to the time domain
and only give them activation variables $m$.
The $e$ current must provide fast positive feedback, so $E_e > 0$ and $\tau_{e,m} < \tau_{i,m}$.
The $i$ current must give slow negative feedback, so $E_i < 0$.
The model is
\begin{equation}
   \begin{aligned}
      \tau_{e,m} \dot{v}_e &= v - v_{e,m} \\
      g_e &= \bar{g}_e \text{relu} \left( v_{e,m} - v_{e,m}^\text{th} \right) \\
      \tau_{i,m} \dot{v}_{i,m} &= v - v_{i,m} \\
      g_i &= \bar{g}_i \text{relu} \left( v_{i,m} - v_{i,m}^\text{th} \right).
   \end{aligned}
   \label{eq:canonical_temporal_excitability}
\end{equation}
together with Equation \ref{eq:CNN_v_dot}.
Here, the $\tau_{x,m}$ are the time constants of the memductances and
$v^\text{th}_{x,m}$ are their thresholds.
Implementing this with the parameters given in Table \ref{tab:canonical_model_temporal_params}
yields a firing threshold analogous to that of biophysical models,
which is demonstrated in Figure \ref{fig:canonical_excitability}.

Due to the model's simplicity,
we can determine the conditions for excitability.
Consider the model starting from rest with $v = v_e = v_i = 0$ 
and $i_\text{app}$ being applied at $t = 0$
with sufficient energy to push $v$ above $v_{e,m}^\text{th}$.
To get positive feedback, we require $d i_\text{tot}/d v < 0$.
As $\tau_{e,m} < \tau_{i,m}$, we have $v_{e,m} \approx v$,
but $v_{i,m}$ remains close to rest: $v_{i,m} \approx 0$.
Then $i_\text{tot} \approx g_l v + \bar{g}_e \left( v - v_{e,m}^\text{th} \right) \left(v - E_e \right)$,
so $d i_\text{tot}/d v \approx g_l + \bar{g}_e \left( 2v - E_e - v_+^\text{th} \right)$,
and the criterion for positive feedback becomes
\begin{equation}
   v_{e,m}^\text{th} < v < \frac{1}{2} \left( - \frac{g_l}{\bar{g}_e} + E_e + v_{e,m}^\text{th} \right).
   \label{eq:pos_conductance_ineq}
\end{equation}

When $v > v_{i,m}^\text{th}$ and both the $e$ and $i$ currents have actived,
we require negative feedback, so $di_\text{tot} / dv > 0$.
The derivative becomes $di_\text{tot} / dv = g_l + 2(\bar{g}_e - \bar{g}_i)v - \bar{g}_e\left(v_{e,m}^\text{th} + E_e\right) - \bar{g}_e\left(v_{i,m}^\text{th} + E_i\right)$ and the criterion for positive differential conductance (negative feedback) with both channels active is
\begin{equation}
   v > \max\left\{ \frac{\bar{g}_e (E_e + v_{e,m}^{\text{th}}) + \bar{g}_i (E_i + v_{i,m}^{\text{th}}) - g_l}{2(\bar{g}_e + \bar{g}_i)}, \, v_{e,m}^{\text{th}}, \, v_{i,m}^{\text{th}} \right\}.
   \label{eq:neg_conductane_ineq}
\end{equation}

The mechanism described above is robust to uncertainty in the parameter values; excitability only requires an $E_e$ sufficiently large so that Equation \ref{eq:pos_conductance_ineq} is satisfied,
and then a $\bar{g}_i > \bar{g}_e$ and sufficiently negative $E_i$ so that Equation \ref{eq:neg_conductane_ineq} is satisfied for all $v > 0$.
This will ensure that there is a region of negative differential conductance (that is, positive feedback)
in the fast timescale and positive differential conductance (negative feedback) overall. Temporal excitability is the result of this mixed-feedback structure,
as Figure \ref{fig:canonical_excitability} shows.

\begin{figure}
   \centering
   \includegraphics[width=\columnwidth]{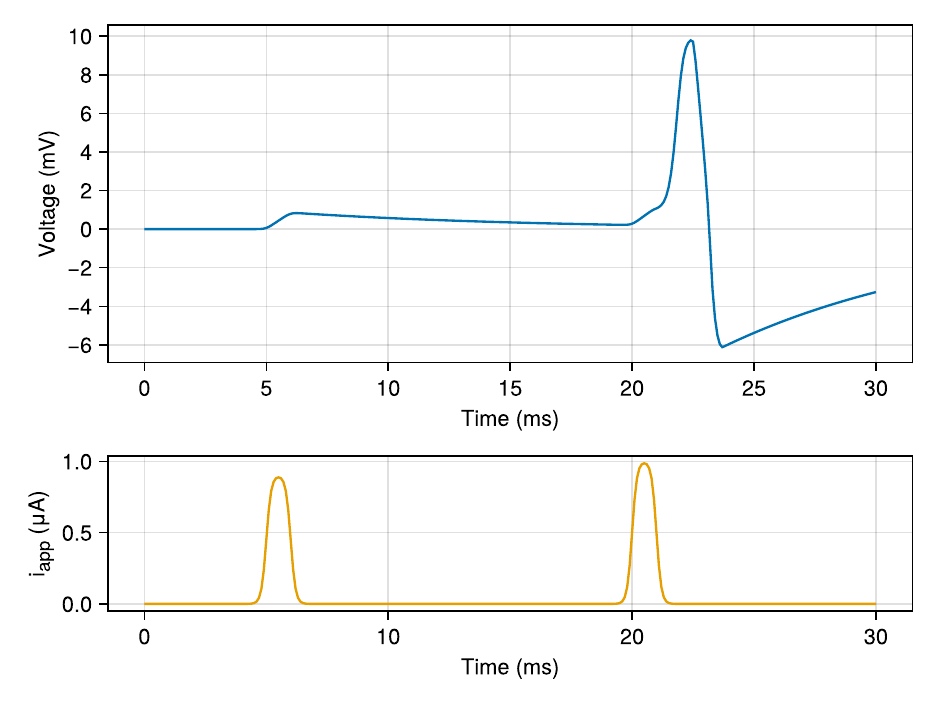}
   \caption{
      Temporal excitability of the memristive neuron model.
      A small increase in $i_\text{app}$ brings the voltage over its threshold,
      resulting in a large pulse-like voltage trajectory (top).
      This threshold behavior is qualitatively equivalent to the excitability of  Hodgkin-Huxley model,
      cf.\ Figure \ref{fig:HH-excitability}.
   }
   \label{fig:canonical_excitability}
\end{figure}

\subsection{A model of  spatial excitability}
\label{subsec:canonical_spatial_excitability}

%\tsjb{
%   \begin{itemize}
 %     \item Point out that Amari's model is not memristive: instead of inventing a kernel as he does, we introduce a spatial conductance.
 %  \end{itemize}
%}
Amari's model exhibits the same mixed-feedback structure as the Hodgkin-Huxley model. However it is not memristive, that is, the feedback currents are not Ohmic. 

To obain a memristive model of spatial excitability, we retain the key properties of Amari's model: we choose as our spatial filter
\begin{equation}
   w(x) = e^{-|x| / \sigma},
\end{equation}
where the parameter $\sigma$ again controls the spatial scale of the interactions.
Because the kernel $w$ now defines a memductance, 
and not a current (as with Amari), it must be nonnegative.

%\tsjb{Make sure to clarify negative conductance as it's used here and negative conductance as in positive feedback, which is mentioned elsewhere}
Therefore, to differentiate between excitation and inhibition,
we separate the neural field into excitatory and inhibitory populations
which we label $E$ and $I$, with membrane voltages $v_E$ and $v_I$.
To examine how this can lead to spatial excitability,
we model $g_e$ and $g_i$ as synaptic currents between the $E$ and $I$ populations,
with their memductances following
\begin{equation}
   \begin{aligned}
      \tau_\text{syn} v_{\text{syn},t} &= v - v_{\text{syn},t} \\
      v_{\text{syn},st} &= \left( v_{\text{syn},t} * w \right) (x, t) \\
      g_\text{syn} (x,t) &= \bar{g}_\text{syn} \text{relu} \left( v_{\text{syn}, st} - v_\text{syn}^\text{th} \right)(x, t), \\
\end{aligned}
\label{eq:canonical_spatial_conductance}
\end{equation}
where the subscript $t$ denotes the temporally filtered $v_\text{syn}$,
and the subscript $s$ denotes the spatially filtered field,
$v^\text{th}_\text{syn}$ is the memductance's threshold,
and $*$ denotes the operation of the spatial convolution.

Each population is connected to itself and to the other,
meaning that the $g_e$ memductance for both $\dot{v}_E$ and $\dot{v}_I$ takes $v_E$ as input,
and analogously $g_i$ takes $v_I$ as input for both populations,
as is depicted in Figure \ref{fig:E-I-diagram}.
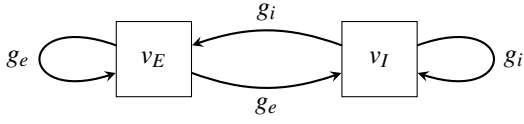
\begin{figure}
    \centering
    \begin{tikzpicture}[node distance=2.5cm, >=stealth]
        % Nodes
        \node (E) [box] at (0,0) {$v_E$};
        \node (I) [box] at (3,0) {$v_I$};

        % Arrows
        % E → E (self-loop on the left)
        \draw[arrow] (E) to[out=160, in=200, looseness=10] node[midway, left] {$g_e$} (E);

        % E → I (straight)
        \draw[arrow] (E) to[bend right=20] node[midway, below] {$g_e$} (I);

        % I → I (self-loop on the right)
        \draw[arrow] (I) to[out=20, in=-20, looseness=10] node[midway, right] {$g_i$} (I);

        % I → E (curved to avoid overlap with E → I)
        \draw[arrow] (I) to[bend right=20] node[midway, above] {$g_i$} (E);

    \end{tikzpicture}
    \caption{Relationship between $v_E$, $v_I$ and $g_e$, $g_i$ to model spatial excitability.}
    \label{fig:E-I-diagram}
\end{figure}

To obtain local excitation with long-range inhibition
the $\sigma$ parameters of the memductance kernels are adjusted for the excitatory and inhibitory synapses,
so that $\sigma_E < \sigma_I$.
Similarly, following physiological data, we choose $\tau_{E, \text{syn}} < \tau_{I, \text{syn}}$,
so that synaptic excitation works faster than synaptic inhibition.

The steady-state voltage response of the $E$ population in the spatial memristive model
with the parameters given in Table \ref{tab:canonical_model_spatial_params}
to two different $i_\text{app}$ currents is shown in Figure \ref{fig:canonical_spatial_excite}.
The first pulse is subthreshold, which elicits a small and brief response.
The second pulse, although only slightly higher in amplitude, is superthreshold. 
As in Figure \ref{fig:amari_spatial_excitability} with Amari's model,
the upshot is a localized persistent bump in activity.
This shows that a simple spatial CNN model of memductances is sufficient to model spatial excitability.

\begin{figure}
   \centering
   \includegraphics[width=\columnwidth]{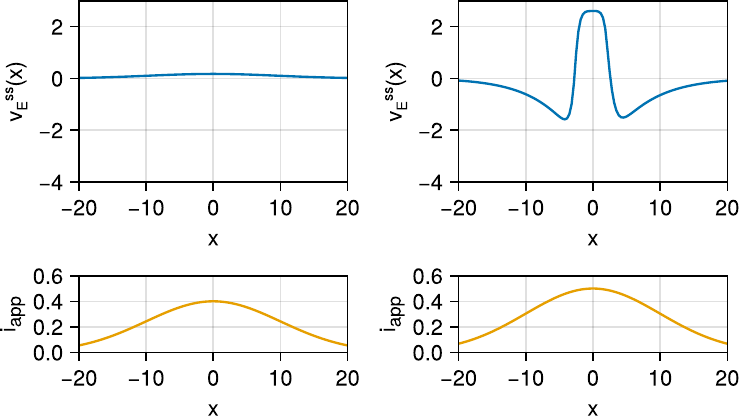}
   \caption{
      Spatial excitability in the CNN model.
      Left: the subthreshold steady-state voltage response $v_E^{ss}(x)$ (top)
      to Gaussian input current (bottom).
      Right: The superthreshold steady-state voltage response $v_E^{ss}(x)$ (top)
      to a Gaussian input current (bottom).
      The first applied pulse does not lead to spatial excitation,
      whereas the second applied pulse does,
      cf.\ Figure \ref{fig:amari_spatial_excitability}.
   }
   \label{fig:canonical_spatial_excite}
\end{figure}

\section{SPATIO-TEMPORAL EXCITABILITY}
\label{sec:space-temp-excite}

%\tsjb{
%\begin{itemize}
 %  \item Message: Spatial network excitability and single-cell temporal excitability are two sides of the same coin
  % \item Putting both together gives one a PING network
%\end{itemize}
%}

To obtain a model of spatio-temporal excitability, we simply combine the temporal mechanism  from Section \ref{subsec:canonical_temporal_excitability} and the spatial mechanism from Section \ref{subsec:canonical_spatial_excitability}. 
The membrane potential of population $k$ evolves according to 
\begin{equation}
   \begin{aligned}
      C\dot{v}_k = -g_l v_k - &\left( g_e(x,t) + g_{e, \text{syn}}(x,t) \right) \left( v_k - E_e \right) \\
      &\left( g_i(x,t) + g_{i, \text{syn}}(x,t) \right) \left( v_k - E_i \right) + i_\text{app},
   \end{aligned}
\end{equation}
where $k \in \{ E , I \}$ indicates the $E$ and $I$ population
and, as before, $g_e$ and $g_i$ are the temporal memductances  with the parameters from Table \ref{tab:canonical_model_temporal_params},
and the dynamics of $g_{e, \text{syn}}$ and $g_{i, \text{syn}}$ are dependent on $v_E$ and $v_I$ respectively
and give the network interactions.

The resulting model unifies the temporally excitable behavior of the Hodgkin-Huxley model
and the spatially excitable behavior of Amari's model (Figure \ref{fig:canonical_spatio-temporal_excitation}).
The purely temporal currents have the biophysical interpretation of {\it internal} memductances,  or ion channels, that is, 
the memductance at position $x$ depends on the voltage at $x$ only. The spatial currents have the
biophysical interpretation of synaptic currents: they obey the same Ohmic law as internal memductances,
with the only difference that their memductance depends on ``pre-synaptic'' voltages, that is, on the voltage
of other neurons in the population. 

\begin{figure}
   \centering
   \includegraphics[width=\columnwidth]{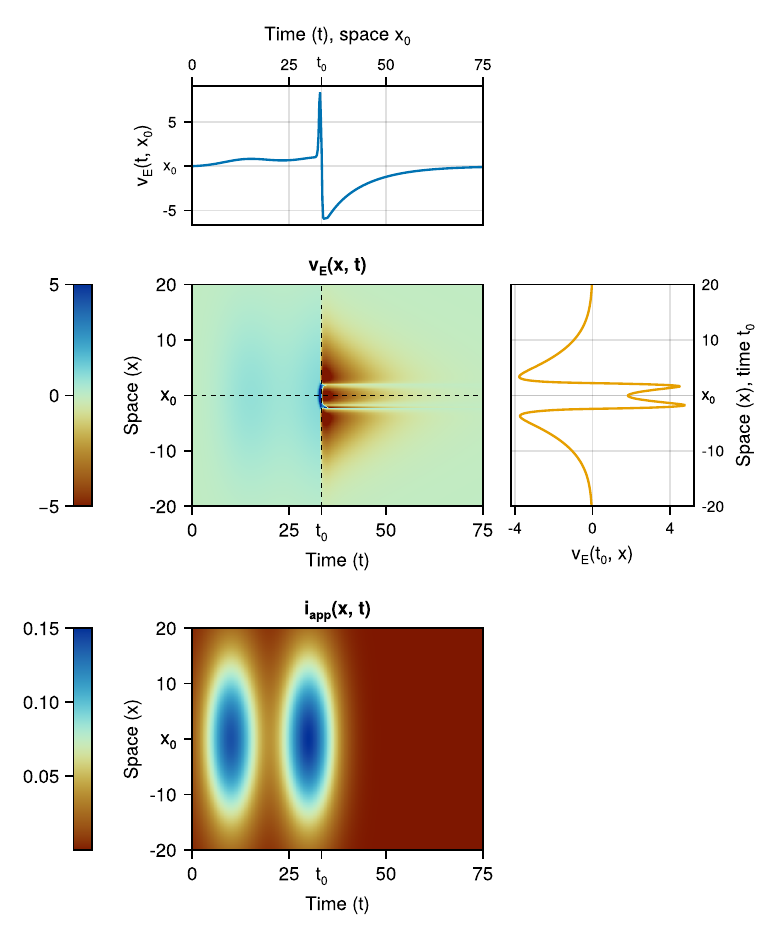}
   \caption{
      Spatio-temporal excitability in the canonical model.
      The response of the excitatory voltage $v_E(x, t)$ (central top figure)
      in response to an applied current $i_\text{app}(x, t)$ (bottom figure).
      The temporal profile of $v_E$ at $x_0 = 0$ is plotted in the upper inset,
      whilst the spatial profile of $v_E$ along $t_0 = 33$ is plotted in the upper-right inset.
      The temporal profile shows temporally excitable behavior (cf. Figure \ref{fig:HH-excitability})
      and the spatial profile shows spatial excitability (cf. Figure \ref{fig:amari_spatial_excitability}).
      }
   \label{fig:canonical_spatio-temporal_excitation}
\end{figure}

\section{CONCLUSIONS AND FUTURE WORK}

\subsection{Conclusions}

Temporal excitability is the result of fast positive feedback tempered by slower negative feedback on the membrane potential.
Spatial excitability is the result of short-ranged excitatory connections
coupled with longer ranged inhibitory connections to limit the excitation and overall activity.
Our paper proposes a methodology that  unifies the modeling principles of those two behaviors.

The architecture of the proposed model retains key properties of the biophysical model of Hodgkin and Huxley: 
a neuron is modeled as an RC circuit controlled by a bank of parallel voltage-gated current sources.
The only part of the model that differentiates its temporal and spatial scales is in the memductances, 
that is, the mapping from the entire neural field $v(x,t)$ to a specific memductance.

The proposed model highlights that the mechanism of excitability is only the result of a mixed-feedback
architecture that is analogous in time and in space. The excitability property is robust to the details
of the memductance model, meaning that excitability can be simulated, analyzed, and designed 
at different scales. 

A key benefit of the proposed model is that it retains the full biophysical interpretability of
traditional biophysical models. Purely temporal memductances have the interpretation 
of ion channels, that is, the voltage dependence is only temporal but not spatial, whereas
spatial currents retain the interpretation of synaptic currents, whose memductance is gated
by other neurons in the population.

\subsection{Future Work}
\label{subsec:future-work}

While preliminary, the results of this paper have sigificant potential to simulate, analyze, 
and design neuronal behaviors across scales. 

The mixed-feedback and biophysical architecture of the model should facilitate
the generalization of the excitability property in this paper to more complex behaviors.
For instance, we expect a smooth generalization from spiking to bursting behaviors
using the multi-scale mixed feedback architecture highlighted for instance in \cite{ribar2021neuromorphic}
and references therein.

We also envision a significant potential for a framework that is algorithmically scalable.
Large-scale simulations of memristive models will rely on the operator-theoretic
framework developed in \cite{chaffey2023monotone} and \cite{shahhosseini2024operator}.
The methodology in those papers is to solve the neural field as the fixed point of an operator in the
space of current-voltage trajectories rather than by integration in time and space. 
The framework makes uses of efficient splitting methods that can leverage the highly parallel structure
of neuronal circuits and their decomposition in simple elements.

\section{APPENDIX}
\label{sec:methods}

For the Hodgkin-Huxley model in Figure \ref{fig:HH-excitability},
the parameters in Table \ref{tab:hh_parameters} were used. 

\begin{table}[h]
\centering
\caption{Hodgkin-Huxley model conductance and reversal potential parameters.}
\begin{tabular}{|c|c|c|c|c|c|}
\hline
$\bar{g}_{\mathrm{Na}}$ & $E_{\mathrm{Na}}$ & $\bar{g}_{\mathrm{K}}$ & $E_{\mathrm{K}}$ & $\bar{g}_{\mathrm{L}}$ & $E_{\mathrm{L}}$ \\
\hline
120 & 50 & 36 & -77 & 0.3 & -54.4 \\
\hline
\end{tabular}
\label{tab:hh_parameters}
\end{table}

For Amari's model and $w$ for Figure \ref{fig:amari_spatial_excitability},
the parameters in Table \ref{tab:amari_model_parameters} were used.
\begin{table}[h]
\centering
\caption{Amari's model parameters.}
\begin{tabular}{|c|c|c|c|}
\hline
$\tau$ & $\sigma_E$ & $\sigma_I$ & $\theta$ \\
\hline
3 & 0.3 & 2 & 0.4 \\
\hline
\end{tabular}
\label{tab:amari_model_parameters}
\end{table}

The parameters of the  memristive neuron model with temporal excitability 
used in Figures \ref{fig:canonical_excitability} and \ref{fig:canonical_spatio-temporal_excitation} are given in Table \ref{tab:canonical_model_temporal_params}.
Whever applicable, $C=1$.
\begin{table}[h]
\centering
\caption{Parameters for temporal excitability in the CNN model.}
\begin{tabular}{|c|c|c|c|c|c|c|c|c|}
\hline
$g_l$ & $\tau_{e,m}$ & $\bar{g}_e$ & $E_e$ & $v_{e,m}^\text{th}$ & $\bar{g}_i$  & $\tau_{i,m}$ & $E_i$ & $v_{i,m}^\text{th}$ \\
\hline
0.1 & 0.1 & 1 & 10 &  1 & 10 & 10 & -10 & 1 \\
\hline
\end{tabular}
\label{tab:canonical_model_temporal_params}
\end{table}

The parameters of the CNN model with spatial excitability,
used in Figures \ref{fig:canonical_spatial_excite} and \ref{fig:canonical_spatio-temporal_excitation} are given in Table \ref{tab:canonical_model_spatial_params}.
\begin{table}[h]
\centering
\caption{Parameters for spatial excitability in the CNN.}
\begin{tabular}{|c|c|c|c|c|c|c|c|c|c|c|}
\hline
$g_l$ & $\sigma^E$ & $\tau_\text{syn}^E$ & $v_E^\text{th}$ & $\bar{g}_\text{syn}^E$ & $E_\text{syn}^E$ & $\sigma^I$ & $\tau_\text{syn}^I$ & $v_I^\text{th}$ & $\bar{g}_\text{syn}^I$ & $E_\text{syn}^I$ \\
\hline
0.1 & 0.5 & 0.1 & 2 & 10 & 10 & 5 & 1 & 2 & 3 & -10 \\
\hline
\end{tabular}
\label{tab:canonical_model_spatial_params}
\end{table}

The applied current followed $i_\text{app}(x,t) = i_{\text{pulse}}^{(1)} + i_{\text{pulse}}^{(2)}$ where
$i_\text{pulse}(x, t) = A \exp\left(-\frac{x^2}{2\sigma_x^2}\right) \exp\left(-\frac{(t - t_0)^2}{2\sigma_t^2}\right)$.
For Figures \ref{fig:amari_spatial_excitability},
\ref{fig:canonical_spatial_excite},
and \ref{fig:canonical_spatio-temporal_excitation},
the $i_\text{app}$ parameters from Table \ref{tab:combined_stimulation_parameters}
were used.

For Figure \ref{fig:amari_spatial_excitability} values $u^{ss}$ were read off at $t=205$ and $t=305$.

For Figure \ref{fig:canonical_spatial_excite} values for $v_E^{ss}$ were read of at $t=100$ and $t=200$.

\begin{table}[ht]
\centering
\caption{Stimulation parameters of $i_\text{app}$ used in Figures \ref{fig:amari_spatial_excitability}, \ref{fig:canonical_spatial_excite}, and \ref{fig:canonical_spatio-temporal_excitation}.}
\begin{tabular}{|c!{\vrule width 1pt}c|c|c|c|c|c|}
\hline
Figure & $A^{(1)}$ & $A^{(2)}$ & $\sigma_x$ & $\sigma_t$ & $t_0^{(1)}$ & $t_0^{(2)}$ \\
\hline
\ref{fig:amari_spatial_excitability} & 2.5 & 2.6 & 5.0 & 5.0 & 10 & 210 \\
\ref{fig:canonical_spatial_excite} & 0.4 & 0.5 & 10 & 5 & 5 & 105 \\
\ref{fig:canonical_spatio-temporal_excitation} & 0.14 & 0.15 & 10 & 5 & 10 & 30 \\
\hline
\end{tabular}
\label{tab:combined_stimulation_parameters}
\end{table}

%%%%%%%%%%%%%%%%%%%%%%%%%%%%%%%%%%%%%%%%%%%%%%%%%%%%%%%%%%%%%%%%%%%%%%%%%%%%%%%%
\section{ACKNOWLEDGMENTS}

This work was supported by funding from the European Research Council under the European Union's Horizon 2020 research and innovation program / ERC Advanced Grant: SpikyControl (no. 7101054323),
and by reviewer's comments, both of which we are grateful for.

%%%%%%%%%%%%%%%%%%%%%%%%%%%%%%%%%%%%%%%%%%%%%%%%%%%%%%%%%%%%%%%%%%%%%%%%%%%%%%%%

\bibliographystyle{ieeetr}
\bibliography{CDC_2025}

\end{document}